\documentclass[review]{elsarticle}
\usepackage{lineno,hyperref}
\usepackage[english]{babel}
\usepackage[numbers]{natbib}
\usepackage{xcolor}
\usepackage{latexsym,amsmath,amssymb,amsbsy,graphicx,geometry}
\modulolinenumbers[5]

\begin{document}
	
	\begin{frontmatter}

\title{Neural network as a tool for design of amorphous metal alloys with desired elastoplastic properties}

\author[kfu]{B.N. Galimzyanov\corref{cor1}}
\cortext[cor1]{Corresponding author}
\ead{bulatgnmail@gmail.com}

\author[kfu]{M.A. Doronina}
\ead{anatolii.mokshin@mail.ru}

\author[kfu]{A.V. Mokshin}
\ead{anatolii.mokshin@mail.ru}

\address[kfu]{Kazan Federal University, 420008 Kazan, Russia}

\begin{abstract}
	The development and implementation of the methods for designing amorphous metal alloys with desired mechanical properties is one of the most promising areas of modern materials science. Here, the machine learning methods appear to be a suitable complement to empirical methods related to the synthesis and testing of amorphous alloys of various compositions. In the present work, it is proposed a method to determine amorphous metal alloys with mechanical properties closest to those required. More than $50\,000$ amorphous alloys of different compositions have been considered, and the Young's modulus $E$ and the yield strength $\sigma_{y}$ have been evaluated for them by the machine learning model trained on the fundamental physical properties of the chemical elements. Statistical treatment of the obtained results reveals that the fundamental physical properties of the chemical element with the largest mass fraction are the most significant factors, whose values correlate with the values of the mechanical properties of the alloys, in which this element is involved. It is shown that the values of the Young's modulus $E$ and the yield strength $\sigma_{y}$ are higher for amorphous alloys based on Cr, Fe, Co, Ni, Nb, Mo and W formed by the addition of semimetals (e.g. Be, B, Al, Sn), nonmetals (e.g. Si and P) and lanthanides (e.g. La and Gd) than for alloys of other compositions. Increasing the number of components in alloy from $2$ to $7$ and changing the mass fraction of chemical elements has no significantly impact on the strength characteristics $E$ and $\sigma_{y}$. Amorphous metal alloys with the most improved mechanical properties have been identified. In particular, such extremely high-strength alloys include Cr$_{80}$B$_{20}$ (among binary), Mo$_{60}$B$_{20}$W$_{20}$ (among ternary) and Cr$_{40}$B$_{20}$Nb$_{10}$Pd$_{10}$Ta$_{10}$Si$_{10}$ (among multicomponent).
\end{abstract}


\end{frontmatter}

\section{Introduction}

Amorphous metal alloys are the promising materials for the automotive, aerospace, energy, electronics and medical technology industries~\cite{Wang_Dong_2004,Chen_2011,Yeh_Huang_2022,Huang_Lin_2023}. High corrosion resistance, high magnetic permeability, superior mechanical strength, high fracture toughness, high elastic strain limit and high formability are just some of the unique set of properties that make amorphous metal alloys widely applicable~\cite{Kruzic_2016,Louzguine-Luzgin_2017,Li_Lu_2019}. Such the combination of properties is directly due to the absence of structural order accompanied by defects that is typical for crystalline analogues~\cite{Malygin_2011,Galimzyanov_Doronina_2021,Anikeev_2022,Galimzyanov_IJSS_2021}. However, despite all the advantages of amorphous metal alloys, their production is complicated by the fact that the formation of a stable disordered structure depends strongly on alloy composition (i.e. number of components, type of added chemical elements) and its preparation protocol (i.e. cooling and compression procedures, initial and final melt temperatures)~\cite{Lesz_2017,Louzguine-Luzgin_2022,Tournier_Ojovan_2022,Ojovan_Tournier_2021,Galimzyanov_JPCM}. 

Amorphous metal alloys are actively studied for more than 80 years, beginning, in particular, with the works of Kramer~\cite{Kramer_1939,Brenner_1950}. One of the first methods of practical formation of alloys with amorphous structure was based on the so-called electrodeposition process. Later, in the 60's of the 20th century, the first works related with formation of amorphous metal films by rapid cooling of the corresponding melts were appeared~\cite{Duwez_1967,Pond_Maddin_1969}. As it turned out later, amorphization of metallic melts of almost any composition is possible if extremely fast cooling is used. The next stage in the development of the amorphous alloy formation methodology concerned the consideration of alloys in eutectics, where it was found that bulk amorphous samples of more than $1$ mm thickness can be formed~\cite{Peker_Johnson_1993}. Further attention in this area focuses on some aspects. Namely, the mechanical properties of bulk amorphous metal alloys are strongly dependent on alloy composition and chemical purity of raw material. The strength properties of amorphous alloys can be significantly reduced due to the presence of impurities. Moreover, bulk amorphous metal alloys are inherently fragile. Therefore, in the early 2000's, studies were aimed to improve the alloy hardening methods as well as to determine the relationship between the key mechanical properties of amorphous metal alloys, which include the Young's modulus $E$, the yield strength $\sigma_y$ and the strength $\sigma_f$~\cite{Demetriou_Kaltenboeck_2009,Conner_Li_2004}. It has been shown that the relationship between the hardness $H$ (by Vickers method), the strength $\sigma_f$, the Young's modulus $E$ and the yield strength $\sigma_y$ of amorphous metal alloys is close to linear and can be reproduced, for example, by Tabor's relation $H=K\sigma_y$, by Johnson's model $H=\sigma_y(a+b\ln[cE/\sigma_{y}])$ and by relation $\sigma_f=dE^{1/2}$ (here, $K$, $a$, $b$, $c$ and $d$ are constants)~\cite{Inoue_Shen_2006,Zhang_Subhash_2006,Yuan_Xi_2011}. These studies found that amorphous metal alloys with large values of $E$ and $\sigma_y$ are characterized by high hardness $H$ and strength $\sigma_f$. 

The synthesis of amorphous metal alloy with desired mechanical properties may require listing various combinations of compositions followed by mechanical testing. This makes the process of synthesizing new alloys extremely difficult and significantly increases the costs. Then, methods of computer design seem to be a suitable support for empirical methods at the stage of determining amorphous metal alloys with desired mechanical properties~\cite{Mokshin_2020,Mokshin_Khabibullin_2022}. In recent decades, rapid development of information technologies as well as automation of data collection and storage processes contribute to accumulation and systematization of information about the physical and mechanical properties of bulk amorphous metal alloys glasses~\cite{White_2013,Rodrigues_Florea_2021,Agrawal_Choudhary_2016,Tripathi_Kumar_2020}. The methods of machine learning operate with large arrays of the data and allow us to determine the relationship between composition and properties of alloys both already known and not previously known~\cite{Shokrollahi_Gu_2022,Merayo_Camacho_2020,Ciupan_Jucan_2018,Galimzyanov_Materials_2023}. For example, Xiong and co-authors have been developed a machine learning model that can predict the glass-forming ability and elastic moduli of bulk metallic glasses based on the fundamental atomic properties, chemical and physical properties obtained from experiments or density functional theory simulations~\cite{Xiong_Shi_2020}. These results find the importance of individual chemical element properties and macroscopic properties in determining the strength characteristics of amorphous alloys. The results obtained by Khakurel et al. established that the average concentration of valence electrons, the atomic radius and the melting temperature are the key properties, which are correlated with the Young's modulus of compositionally complex alloys~\cite{Khakurel_Taufique_2021}. The results of this work can also be extended to amorphous metal alloys, as it is confirmed in Refs.~\cite{Amigo_Palominos_2023,Xiong_Zhang_2019}. In addition, as it was found in Ref.~\cite{Galimzyanov_PhysicaA_2023} using a machine learning model, the Young's modulus of metal alloys under normal conditions correlates with the yield strength and with the glass transition temperature. In this case, the specificity of ``chemical formula'' of alloy, which is determined by the molar mass and the number of components, is not as important as is usually expected. Johnson and Samwer have found that the mechanical properties (elastic constants, compressive yield strength, elastic strain limit) of $30$ bulk metallic glasses as functions of the scaled temperature $T_{R}/T_{g}$ obey the universal law $\propto a-b(T_{R}/T_{g})^{2/3}$, where $a$ and $b$ are the constants, $T_R$ is the room temperature, $T_g$ is the glass transition temperature~\cite{Johnson_Samwer_2005}. The results of this work systematize existing knowledge about the mechanical properties of amorphous alloys. An artificial neural network has created by Jeon and co-authors for designing Fe-based amorphous metal alloys with the desired crystallization temperature and glass transition temperature~\cite{Jeon_Seo_2021}. Thus, all these studies show that the machine learning methods are suitable tool to find new amorphous alloys with required physical and mechanical properties. Despite the significant number of such studies, little attention has been paid to the development of methods for determining previously unknown amorphous alloys with the desired mechanical properties.

The present work proposes a new method for determining amorphous metal alloys of arbitrary composition based on a large set of empirical data. The originality of this method is that it is based on a machine learning model capable of predicting the Young's modulus and the yield strength of amorphous alloys taking into account the fundamental properties of each chemical element that forms the alloys. It is quite significant that the obtained results lead to new knowledge, which will contribute to the determination of amorphous metal alloys that maximally satisfy the required mechanical properties.

\section{Method for determining the mechanical properties of amorphous metal alloys}

\subsection{General strategy of the method}

The developed method for determining amorphous metal alloys is based on a machine learning model, which is an artificial neural network of direct propagation. The main advantage of this method is the possibility to calculate the Young's modulus $E$ and the yield strength $\sigma_{y}$ both for known amorphous metal alloys and for alloys that are yet to be synthesized. The developed method makes it possible to determine $E$ and $\sigma_{y}$ of alloys, whose number of components varies in the range from $2$ to $7$. Note that such the number of components is ordinary for the majority of known metal alloys. In addition, the proposed method can be adapted to identify alloys with large number of components at the presence of appropriate data for neural network training. The composition and mass fraction of chemical elements in the generated alloys are the control parameters, which allow us to construct a diverse set of compounds.

\begin{figure}[ht!]
	\centering
	\includegraphics[width=14cm]{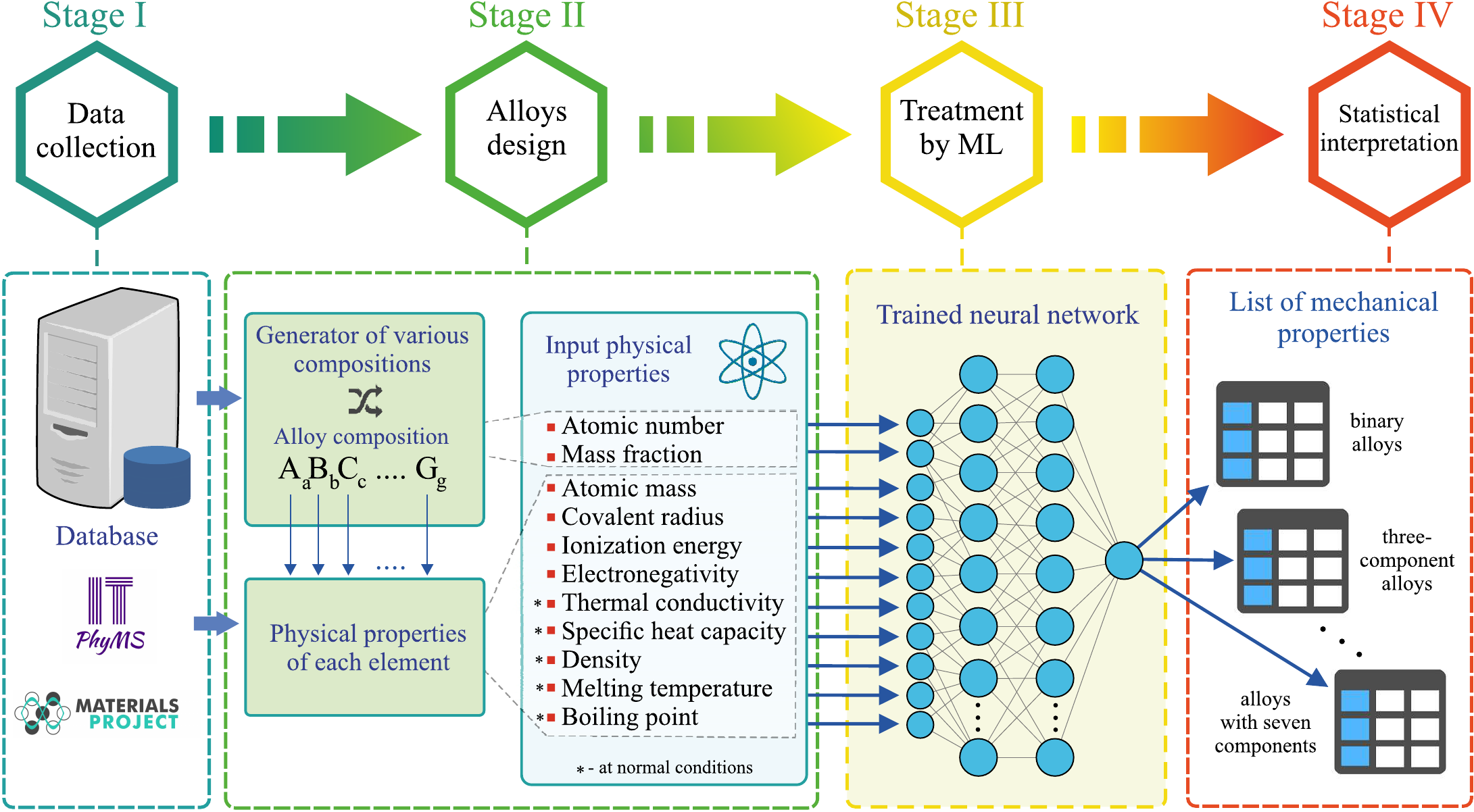}
	\caption{Four-stage scheme of the method for determining amorphous metal alloys and calculating their mechanical properties.}
	\label{fig_1}
\end{figure}  

The general strategy for determining amorphous metal alloys implemented in this work consists of four main stages [see Figure~\ref{fig_1}]:
\begin{itemize}
\item \noindent Stage I. This stage includes the process of data collection and systematization of information about the properties of multicomponent amorphous metal alloys based on Al, Au, Ca, Co, Cu, Fe, La, Hf, Mg, Ni, Pd, Pt, Sc, Ti, W, Zr, etc., as well as information about the properties of the other additional chemical elements involved in the formation of these alloys. Among these properties are the atomic mass $m_{a}$, the covalent radius $r_c$, the ionization energy $E_{i}$ and the electronegativity $\chi$, which characterize the nature of the chemical element [see Table~\ref{tab_1}]. This choice is due to the following reasons. First, these parameters most clearly define the possible physical and chemical bonds between the elements, which can either promote or inhibit the formation of an amorphous structure. For example, according to the empirical rule proposed by Inone et al. in the early 1990's~\cite{Inone_1998}, the difference in atomic sizes must be greater than $12$~\% for good amorphization of a liquid. Secondly, most of the intrinsic properties of chemical elements (especially of the same type) are correlated. In addition, the thermal conductivity $\lambda$, the specific heat capacity $C_{s}$, the density $\rho$, the melting temperature $T_{m}$ and the boiling temperature $T_{b}$ of chemical elements at normal conditions are used. The atomic number $Z$ and the mass fraction $m_{f}$ of each chemical element in the alloy are used to characterize the alloy composition. The Young's modulus $E$ and the yield strength $\sigma_{y}$ are also applied, whose values are known for the considered amorphous alloys. The values of all the listed physical properties are taken from the database {\it ITPhyMS} (Information technologies in physical materials science)~\cite{st_itphyms} and the database {\it Materials Project}~\cite{st_matproject} as well as from Refs.~\cite{Galimzyanov_Materials_2023,Wang_2006,Wang_2012,Chen_Li_2011,Qu_Liu_2015} [see Supplementary data of the present work]. These properties are characterized by different physical dimensions and by different ranges of values. Therefore, the properties are calibrated so that their values vary in the range [$0$; $1$]. The calibration is done according to the rule 
\begin{equation}\label{eq_normalization}
\text{Property}'=\frac{\text{Property}-\text{Value}_{\text{min}}}{\text{Value}_{\text{max}}-\text{Value}_{\text{min}}},
\end{equation}
where ``$\text{Value}_{\text{min}}$'' and ``$\text{Value}_{\text{max}}$'' are the smallest and largest known values of the ``$\text{Property}$''. Moreover, all these listed properties correlate with the mechanical properties of materials. For example, Xiong et al. have shown that the accuracy of predicting the mechanical properties of amorphous metal alloys is improved when the quantities $m_{a}$, $Z$, $r_c$, $\rho$, $\lambda$, $T_{m}$ and $T_{b}$ are considered in a machine learning model~\cite{Xiong_Shi_2020}. In addition, the results obtained by Wang based on the analysis of a large set of empirical data for amorphous alloys allow one to establish the existence of correlation between elastic moduli (i.e. Young's modulus, shear modulus, bulk modulus), microstructural features, rheological properties, the glass transition temperature, the melting temperature and the boson peak~\cite{Wang_2006,Wang_2012}.

\item \noindent Stage II. Alloys with different compositions are generated. Taking into account the number of possible components, combinations of all chemical elements and their mass fraction, up to $10^{18}$ different compositions can be determined simultaneously. When obtaining alloys, those chemical elements are selected that are included on the alloys in the training dataset. In the present work, $32$ chemical elements were used including transition metals (Fe, Co, Ni, Cu, etc.), semimetals (B, Al, Sn, etc.), lanthanides (La, Gd, Er, etc.) and alkali and alkaline earth metals (Li, Be, Mg, Ca, etc.). A list of all the considered chemical elements is given on Table S1 in Supplementary data. The mass fraction of the chemical elements in a generated alloy is also set randomly so that the total mass fraction of all chemical elements is equal to $100$~\%. A set of physical properties is created for each chemical element [see Table~\ref{tab_1}].

\item \noindent Stage III. Information about the alloy composition and the physical properties of all the chemical elements is processed by the pre-trained neural network. This neural network evaluates the Young's modulus $E$ and the yield strength $\sigma_{y}$ for all generated alloys. The training procedure of the neural network is discussed in more details in the subsection ``Machine learning model: structure and training''. 

\item \noindent Stage IV. Statistical interpretation of machine learning results is performed.
\end{itemize}
Thus, the proposed method makes it possible to perform a complete cycle of alloy design and determine its mechanical properties: from obtaining the correct alloy composition to calculating the correct values of $E$ and $\sigma_{y}$.
\begin{table}[tbh]
	\centering
	\caption{Physical properties of chemical elements used as input parameters in the artificial neural network.}	
	\begin{tabular}{lcc}
		\hline		\hline
		Property & Symbol & Unit \\
		\hline
		Atomic number      		& $Z$ 		&  -- \\
		Mass fraction of elements    & $m_{f}$ 		&  \% \\
		\hline
		Atomic mass 			& $m_{a}$ 	&  a.e.m. \\
		Covalent radius 		& $r_{c}$ 	&  pm \\
		Ionization energy 		& $E_{i}$ 	&  eV \\
		Electronegativity 		& $\chi$ 	&  -- \\
		Thermal conductivity 	& $\lambda$ &  W/(m$\cdot$K)\\
		Specific heat capacity 	& $C_{s}$ 	&  J/(g$\cdot$K)\\
		Density					& $\rho$ 	&  g/cm$^3$\\
		Melting temperature 	& $T_{m}$ 	&  K \\
		Boiling temperature 	& $T_{b}$	&  K \\
		\hline		\hline
	\end{tabular}\label{tab_1}
\end{table}

\subsection{Machine learning model: structure and training}

The machine learning model is the four-layer artificial neural network. The first layer has $77$ input neurons for the values of $11$ physical properties for all chemical elements of the obtained alloy ($7$ input neurons are allocated to each property because the maximal number of components in the alloy is also seven). If the number of components in the alloy is less than $7$, then the remaining neurons are unused. The next two layers are hidden. The first hidden layer consists of $80$ neurons, while the second hidden layer has $10$ neurons. Note that the number of neurons in the hidden layers is optional. The neural network produces close results with $80$ to $100$ neurons in the first hidden layer and with $10$ to $80$ neurons in the second hidden layer. The fourth layer consists of one neuron that determines the Young's modulus $E$ or the yield strength $\sigma_{y}$. It is important to note that two separate independent neural networks with the same structure are used to calculate the values of $E$ and $\sigma_{y}$.

Calculation of the values of all neurons is carried out by expression~\cite{Chumachenko_Gabbouj_2022}:
\begin{equation}\label{eq_ns_3}
n_{i}^{(k)}=f\left(\sum_{j=1}^{N_{k-1}}w_{ij}^{(k-1)}n_{j}^{(k-1)}+b_{i}^{(k)}\right).
\end{equation}
Here, $n_{i}^{(k)}$ is the value of the $i$th neuron in the $k$th layer ($k=2,\,3,\,4$); $w_{ij}^{(k-1)}$ is the value of the $(k-1)$th layer weight going from a neuron with index $j$ to a neuron with index $i$ from the $k$th layer; $b_{i}^{(k)}$ is the bias weight acting on a neuron with index $i$; $N_{k-1}$ is the number of neurons in the $(k-1)$th layer. The sigmoid $f(x)=1/(1-\exp[-x])$ is applied as the activation function~\cite{Sharma_Athaiya_2020}. 
\begin{figure}[ht!]
	\centering
	\includegraphics[width=15cm]{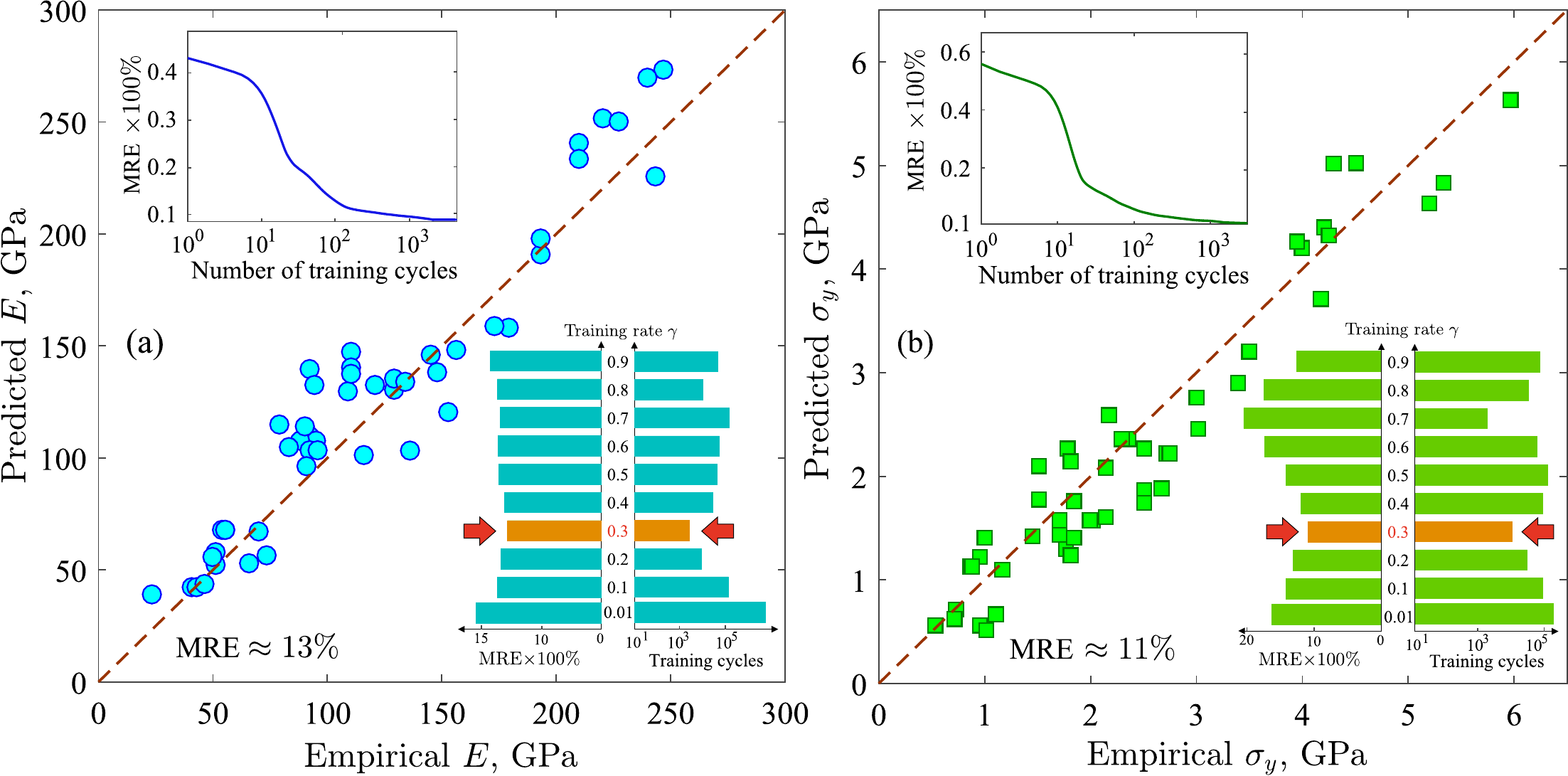}
	\caption{(a) Plot of the predicted Young's modulus $E$ versus the empirical $E$. (b) Plot of the predicted yield strength $\sigma_{y}$ versus the empirical $\sigma_{y}$. Top insets: mean relative error as function of the number of training cycles for $E$ and $\sigma_{y}$. Bottom insets: dependence of the mean relative error and the training cycles on the training rate $\gamma$, from which the optimal value of $\gamma$ (indicated by the red arrows) was determined.}
	\label{fig_2}
\end{figure}  

The neural network is trained using the backpropagation algorithm~\cite{Haykin_2009,Li_Huang_2012}. The values of the weight coefficients are adjusted as follows: 
\begin{equation}\label{eq_leanr_1}
w_{ij}^{(k),\,new}=w_{ij}^{(k)}-\gamma\frac{\partial\xi}{\partial w_{ij}^{(k)}},
\end{equation}
where $\xi$ is the squared error between the output neuron and the desired value of the mechanical property; $\gamma$ is the training rate. In the present work, the training rate is $\gamma=0.3$, which is optimal for the created neural network. At the training rate $\gamma=0.3$, the machine learning model gives the best result for $E$ and $\sigma_{y}$ with the lower MRE at the relatively small number of training cycles [see insets on Figures~\ref{fig_2}(a) and~\ref{fig_2}(b)]. The original dataset is divided into training and validation subsets in proportion 80:20. The training subset consists of amorphous metal alloys based on Al, Au, Ca, Co, Cu, Fe, La, Hf, etc. with different compositions, for which the values of $E$ and $\sigma_{y}$ are known~\cite{Galimzyanov_Materials_2023,Wang_2006,Wang_2012,Chen_Li_2011,Qu_Liu_2015}. The physical properties of the chemical elements of these alloys are also used in the training procedure [see Supplementary data]. To verify the correctness of the machine learning results, the validation subset is applied, which includes amorphous alloys that were not included on the training subset. The criterion to stop the training procedure is the minimal error between the results of the output neuron and the required values from the validation subset.

\subsection{Validation of the machine learning model}

Typically, RMSRE (Root Mean Squared Relative Error), RMSE (Root Mean Square Error), RRMSE (Relative Root Mean Square Error), MSE (Mean Square Error), MAE (Mean Absolute Error) or MRE (Mean Relative Error) are used as indicators for measuring accuracy of results~\cite{Yan_Cui_2021,Sobhanifar_Azizi_2015,Ster_2012,Turco_Mateus_2021, Marques_Ribeiro_2020}. In the present work, it was important to use an indicator that does not depend on units of physical quantities. At the same time, this indicator must be easy to estimate. Therefore, we chose the MRE, which is calculated by expression:
\begin{equation}\label{eq_mre}
\text{MRE}=\frac{1}{N}\sum_{i=1}^{N}\frac{|\mathcal{M}_{\text{ANN}}-\mathcal{M}_{\text{req}}|}{\mathcal{M}_{\text{req}}}\times 100\%.
\end{equation}
Here, $\mathcal{M}=\{E$ or $\sigma_{y}\}$ denotes the mechanical property; $\mathcal{M}_{\text{ANN}}$ is the result of the neural network; $\mathcal{M}_{\text{req}}$ is the required value of the mechanical property; $N$ is the number of items in the validation subset. We find that the MRE is $\sim13$~\% for the Young's modulus $E$ and $\sim11$~\% for the yield strength $\sigma_{y}$. This relatively low MRE indicates a correlation between the predicted and empirical values of the mechanical properties that is also confirmed by the results presented in Figures~\ref{fig_2}(a) and~\ref{fig_2}(b). Moreover, the values of the MRE are stable. This is confirmed by the computed loss functions [see insets on Figures~\ref{fig_2}(a) and~\ref{fig_2}(b)], which reach a plateau after $3\times10^{3}$ training cycles. Thus, the results of the machine learning model are reliable and predictable.

\section{Properties importance scores}

\begin{figure}[ht!]
	\centering
	\includegraphics[width=15cm]{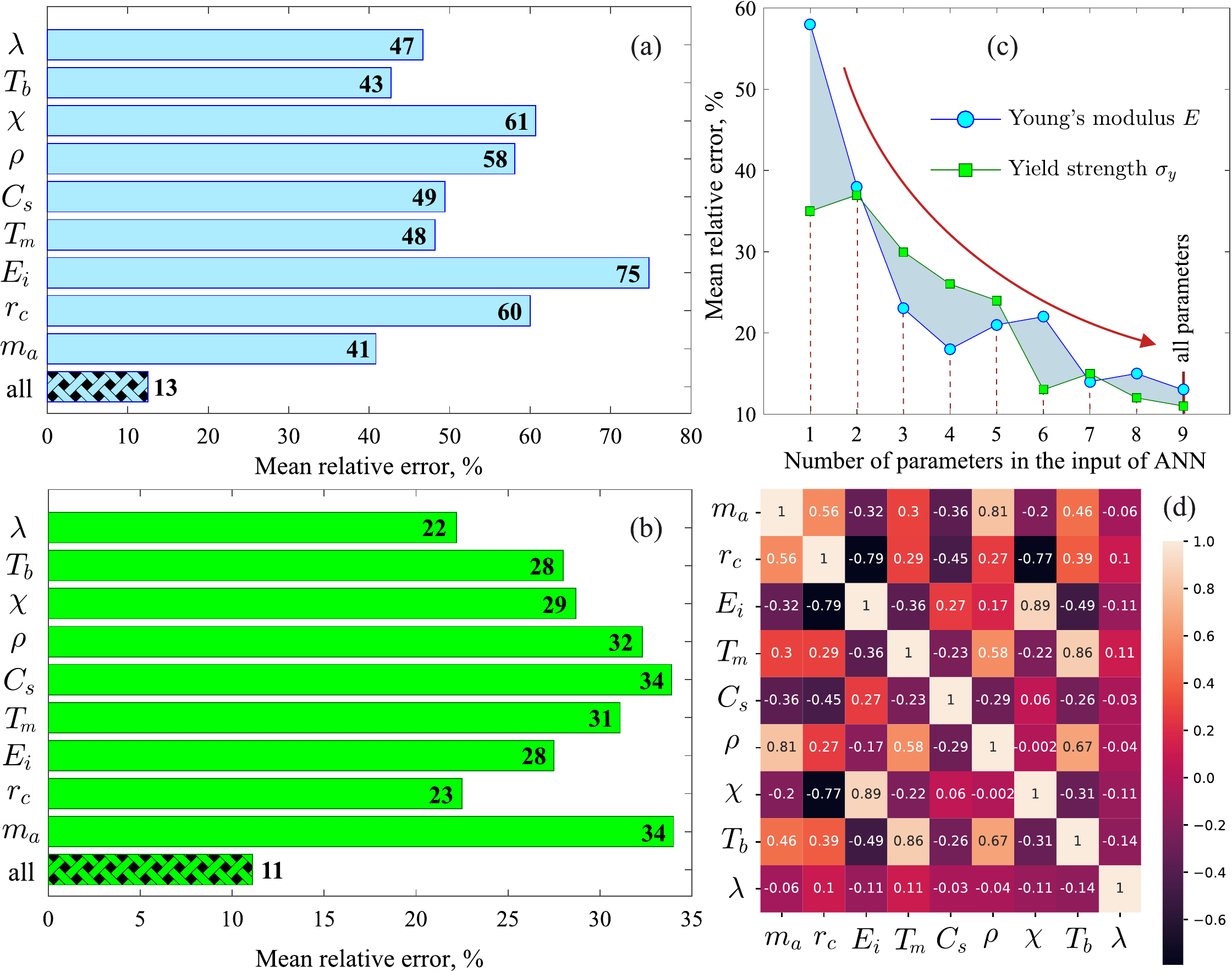}
	\caption{Importance scores for each physical property: (a) in the case of the Young's modulus and (b) in the case of the yield strength. (c) Mean relative error as a function of the number of parameters in the artificial neural network input. (d) Pearson correlation heat map for the considered physical properties.}
	\label{fig_3}
\end{figure}

The analysis of the importance scores shows that all the considered physical properties ($\lambda$, $T_{b}$, $\chi$, $\rho$, $C_{s}$, $T_{m}$, $E_{i}$, $r_{c}$ and $m_{a}$) are necessary for the correct evaluation of the Young's modulus $E$ and the yield strength $\sigma_{y}$ by the machine learning model. As can be seen from Figure~\ref{fig_3}(a), in the case of the Young's modulus $E$, the importance scores of the properties $\lambda$, $T_{b}$, $C_{s}$, $T_{m}$ and $m_{a}$ are similar (MRE$\sim45$~\%), and these physical properties can be recognized as significant factors. The lowest importance scores are observed for the density $\rho$ and for the properties characterizing the chemical nature of the atoms: the ionization energy $E_{i}$ (MRE$\sim75$~\%), the electronegativity $\chi$ (MRE$\sim61$~\%) and the covalent radius $r_c$ (MRE$\sim60$~\%). These properties have less impact on the result of the machine learning model. Furthermore, considering all the properties reduces the error to MRE$\sim13$~\%. The importance scores of the properties in the case of the yield strength $\sigma_{y}$ differ significantly from the Young's modulus $E$ [see Figure~\ref{fig_3}(b)]. The main factors affecting the yield strength $\sigma_{y}$ are the thermal conductivity $\lambda$ (MRE$\sim22$~\%) and the covalent radius $r_{c}$ (MRE$\sim23$~\%). MRE for other parameters is above $28$~\%. Together, these properties lead to the best result, where the error is MRE$\sim11$~\%.

Thus, the formation of the machine learning models for the Young's modulus $E$ and the yield strength $\sigma_y$ using a single physical property ($\lambda$, or $T_{b}$, or $\chi$, or $\rho$, or $C_{s}$, or $T_{m}$,\,$\dots$) produces an error that is much larger than the error when these machine learning models are formed with the entire set of physical properties. Mathematically, such the situation is possible when the correlation between the parameter $E$ (or $\sigma_y$) and some individual physical property appears indirectly (not explicitly). In turn, this means that it is not possible to obtain analytical expressions relating a mechanical property with any parameter of the set ($\lambda$, $T_{b}$, $\chi$, $\rho$, $C_{s}$, $T_{m}$, $E_{i}$, $r_{c}$ and $m_{a}$) and correctly reproducing the results for an arbitrary metal alloy. The methodology of artificial neural networks used in this study makes it possible to obtain a correspondence between the Young's modulus $E$ (or the yield strength $\sigma_y$) and the whole set of the considered physical properties; and this correspondence is reproduced not by an analytical expression, but by the internal structure of the formed neural network. In fact, this feature of this methodology is an advantage when dealing with a fairly large set of parameters.

An additional evaluation of the accuracy of the machine learning model was performed by computing the MRE for different numbers of physical properties in the input of the neural network. Figure~\ref{fig_3}(c) shows that at adding properties in the order of $\rho$, $T_{m}$, $T_{b}$, $\lambda$, $C_{s}$, $m_{a}$, $r_{c}$, $\chi$ and $E_{i}$ the MRE decreases from $\sim58$~\% to $\sim13$~\% for $E$ and from $\sim33$~\% to $\sim11$~\% for $\sigma_{y}$. A rapid decrease of the error is observed when the temperatures $T_{m}$ and $T_{b}$ as well as the quantities $E_{i}$ and $\chi$ have been added, which may be due to their multicollinearity. Figure~\ref{fig_3}(d) shows that the Pearson correlation coefficients for the considered properties take both positive and negative values in the range from $-1$ to $1$~\cite{Jetly_Chaudhury_2021}. For example, the positive correlation between the temperatures $T_{m}$ and $T_{b}$ is due to the fact that an increase in the melting temperature leads to an increase in the boiling temperature~\cite{Malyshev_Makasheva_2014}. An increase in the atomic mass $m_{a}$ of the alloy components usually leads to an increase in its density $\rho$, which leads to a positive correlation between $m_{a}$ and $\rho$~\cite{Kanematsu_Nakao_2016,Karpechev_Efthymiopoulos_2014}. The presence of the pronounced negative correlation between the pairs $r_{c}$, $E_{i}$ and $r_{c}$, $\chi$ is due to the fact that a decrease in the covalent radius $r_{c}$ leads to an increase $E_{i}$ and $\chi$ by increasing the electron density in the atom~\cite{Agmon_2014,Matovi_Yano_2013}.

\section{Statistical interpretation of the results}

In the present study, $50\,000$ different amorphous metal alloys were obtained by the proposed method. All alloys were sorted according to the atomic number $Z$ of the basic chemical element and the number of components in the alloy. Using the trained machine learning model, the values of $E$ and $\sigma_y$ were calculated for each alloy. Then, the average value of the mechanical property was found for each $X$-based alloy consisting $n$ components (where $n=2,\,3,...,\,7$). Here, $X$ denotes a chemical element, the mass fraction of which in the alloy is greater than that of other elements. For example, Al-based binary alloys Al$_{90}$Fe$_{10}$, Al$_{80}$Cu$_{20}$, Al$_{60}$Ni$_{40}$, etc. were selected and the average values of $E$ and $\sigma_y$ were determined for all these alloys. Similar calculations were performed for alloys based on other metals with different number of components. Then, the dependence of the average values of $E$ and $\sigma_y$ on the atomic number $Z$ of the basic chemical element has been determined.

\begin{figure}[ht!]
	\centering
	\includegraphics[width=15cm]{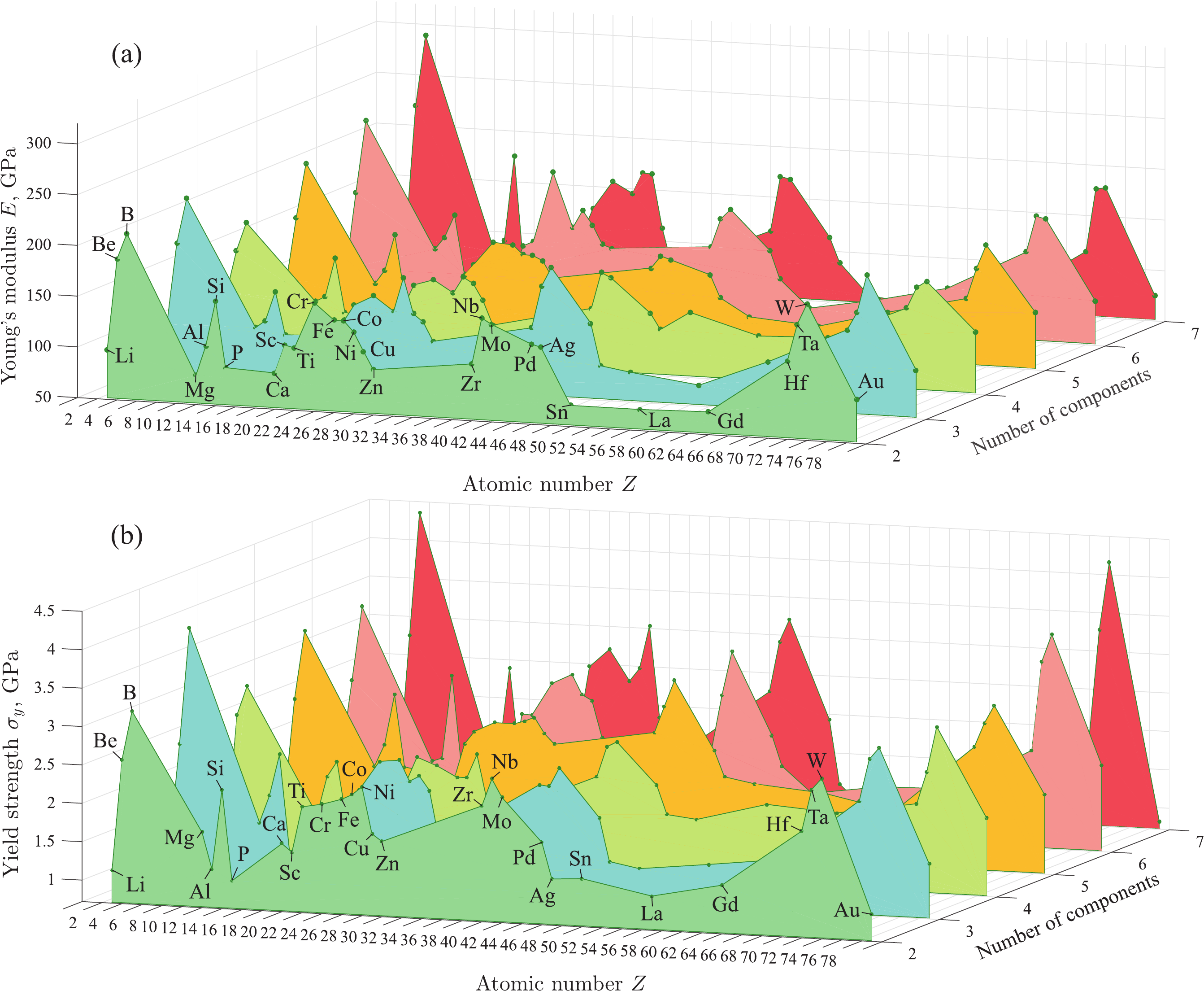}
	\caption{3D plot of the dependence of the mechanical properties on the atomic number $Z$ of the basic chemical element in the alloy and on its number of components: (a) for the Young's modulus $E$ and (b) for the yield strength $\sigma_{y}$.}
	\label{fig_4}
\end{figure}  

In the statistical interpretation, the results reveal that $E$ and $\sigma_{y}$ depend mainly on the properties of the chemical element with the largest mass fraction. As seen in Figure~\ref{fig_4}, changing the number of components in alloy has no significant effect on values of $E$ and $\sigma_{y}$. In the array of $50\,000$ different alloys obtained by the machine learning model, some alloys with the highest Young's modulus $E$ and the yield strength $\sigma_y$ were selected [see Table~\ref{tab_2}]. The results show that these alloys are mainly based on Ti, Cr, Fe, Co, Ni, Zr, Nb, Mo, Pd, Ta and W. For example, these are Mo$_{60}$B$_{20}$W$_{20}$, Co$_{40}$B$_{20}$Be$_{20}$Al$_{20}$, Cr$_{40}$B$_{20}$Nb$_{10}$Pd$_{10}$Ta$_{10}$Si$_{10}$ and Cr$_{30}$Mo$_{20}$W$_{20}$Pd$_{10}$Gd$_{10}$B$_{10}$ alloys for which the mechanical properties are $E>300$~GPa and $\sigma_y>5.0$~GPa. It is important to note that the results for alloys of these compositions were not previously known, although alloys of some related compositions have been studied. So, for example, for W$_{46}$Ru$_{37}$B$_{17}$, Co$_{43}$B$_{31.5}$Fe$_{20}$Ta$_{5.5}$ and Co$_{60}$B$_{35}$Ta$_{5}$ it was experimentally established that $E>250$~GPa and $\sigma_y>5.0$~GPa~\cite{Qu_Liu_2015,Galimzyanov_PhysicaA_2023}. Obtained results reveal that the alloys based on Cr, Mo and W from the group VI-B of the Periodic Table of the Elements have improved mechanical properties. Such the metals as Cr, Mo and W are refractory and have very high hardness~\cite{Braithwaite_Haber_1994,Barnhart_1997,Rieth_Dudarev_2013}. Then, their significant presence in an alloy improves its strength. Note that this fact is also known in metallurgy, where these metals are widely used to increase the hardness of steel alloys, to increase wear resistance and to form wear-resistant coatings (e.g. alloys Cr-Co, Cr-Fe, Mo-Fe, Mo-Cr-Fe, W-Fe, W-Ni-Co)~\cite{Yamanaka_Mori_2014,Baldinozzi_Pontikis_2022}. The machine learning model predicts improved mechanical properties in the case of alloys based on Ti, Cr, Fe, Co, Ni, Zr, Nb, Mo, etc. when these alloys are doped with other metals (e.g. Be, B, Hf), nonmetals (e.g. Si, P) and lanthanides (e.g. La, Gd). The mechanical properties of alloys based on Al, Mg, Ca, Cu, Zn, Ag, Au, Hf, lanthanides, etc. are inferior to those of the alloys based on Ti, Cr, Fe, Co, etc. It should be noted that the relatively high values of $E$ and $\sigma_{y}$ purely for B and Si are due to statistical error, since B-based and Si-based alloys were not used in the training stage of the machine learning model. However, alloys containing B and Si were considered. 

\begin{table}[tbh]
	\scriptsize
	\centering
	\caption{Young's modulus $E$ and yield strength $\sigma_y$ predicted by machine learning model for different amorphous metal alloys. Here, the average accuracy is $\sim88$~\%.}
	\begin{tabular}{clclc}
		\hline		\hline
		Number of components & Alloy & $E$, GPa & Alloy & $\sigma_{y}$, GPa \\
		\hline
		2 & Cr$_{80}$B$_{20}$ & $305$ & Pd$_{60}$B$_{40}$ & $5.13$ \\
		  & W$_{60}$Hf$_{40}$ & $271$ & W$_{60}$Hf$_{40}$ & $5.04$ \\
		\hline
		3 & Mo$_{60}$B$_{20}$W$_{20}$ & $319$ & Mo$_{60}$B$_{20}$Si$_{20}$ & $5.35$ \\
		 & Nb$_{40}$Hf$_{40}$B$_{20}$ & $289$ & Ni$_{60}$B$_{20}$Sc$_{20}$ & $5.17$ \\
		 & Ni$_{60}$B$_{20}$W$_{20}$ & $280$ & Pd$_{60}$B$_{20}$P$_{20}$ & $4.31$ \\
		\hline
		4 & Ag$_{40}$B$_{20}$Sc$_{20}$Ta$_{20}$ & $302$ & Co$_{40}$B$_{20}$Be$_{20}$Al$_{20}$ & $5.27$ \\
 		& Zr$_{40}$Ni$_{20}$B$_{20}$Be$_{20}$ & $285$ & Nb$_{40}$W$_{20}$La$_{20}$B$_{20}$ & $4.88$ \\
		 & Cr$_{40}$B$_{20}$Zr$_{20}$Hf$_{20}$ & $271$ & Ti$_{40}$W$_{20}$Pd$_{20}$B$_{20}$ & $4.70$ \\
		\hline
		5 & Ti$_{30}$Fe$_{30}$B$_{20}$Sn$_{10}$Be$_{10}$ & $296$ & Co$_{40}$B$_{30}$Ag$_{10}$Gd$_{10}$Si$_{10}$ & $5.54$ \\
		 & Pd$_{40}$B$_{20}$Si$_{20}$P$_{10}$Hf$_{10}$ & $289$ & Fe$_{50}$B$_{20}$Mo$_{10}$Ta$_{10}$Ag$_{10}$ & $5.39$ \\
		\hline
		6 & Cr$_{40}$B$_{20}$Nb$_{10}$Pd$_{10}$Ta$_{10}$Si$_{10}$ & $310$ & Cr$_{30}$Mo$_{20}$W$_{20}$Pd$_{10}$Gd$_{10}$B$_{10}$ & $5.62$ \\
		 & Pd$_{40}$Be$_{20}$Mo$_{10}$Ti$_{10}$B$_{10}$Fe$_{10}$ & $306$ & Mo$_{40}$W$_{20}$Pd$_{10}$Gd$_{10}$B$_{10}$Cr$_{10}$ & $5.53$ \\
		 & W$_{30}$B$_{20}$Au$_{20}$Be$_{10}$Nb$_{10}$Ag$_{10}$ & $296$ & Ta$_{20}$Nb$_{20}$Al$_{20}$Au$_{20}$B$_{10}$W$_{10}$ & $5.06$ \\
		\hline
		7 & W$_{20}$Co$_{20}$Nb$_{20}$Ag$_{10}$B$_{10}$Be$_{10}$Mg$_{10}$ & $284$ & W$_{20}$B$_{20}$Ag$_{20}$Nb$_{10}$Si$_{10}$Co$_{10}$Pd$_{10}$ & $5.24$ \\
		 & Cr$_{20}$Ag$_{20}$Ti$_{20}$B$_{10}$Gd$_{10}$Be$_{10}$Mg$_{10}$ & $234$ & Cr$_{20}$Fe$_{20}$W$_{20}$Ca$_{10}$B$_{10}$Sn$_{10}$Be$_{10}$ & $3.78$ \\
		\hline		\hline
	\end{tabular}\label{tab_2}
\end{table}

In Table~\ref{tab_3}, we list $10$ binary and ternary amorphous metal alloys selected from $50\,000$ alloys considered in this study. By simple comparison one can reveal that the predicted $E$ is mainly correlated with the mechanical properties of chemical element with the highest mass fraction. For example, for amorphous Cr$_{80}$B$_{20}$, we find $E\approx305$~GPa, while the Young's modulus of pure crystalline Cr is $E\approx279$~GPa. In the case of amorphous W$_{40}$Mo$_{40}$B$_{20}$, we have $E\approx318$~GPa, while $E$ is $\sim410$~GPa for pure crystalline W. The mechanical properties can vary depending on the concentration of the doped chemical elements and on the class to which these elements belong (metals, nonmetals, lanthanides, etc.). For example, the predicted Young's modulus for Ni$_{40}$Cr$_{40}$Co$_{20}$ is $E\approx58$~GPa. At the same time, the presence of Zr and Si in the Ni-based alloy doubles the Young’s modulus $E$ (i.e. one has $E\approx108$~GPa for Ni$_{40}$Zr$_{40}$Si$_{20}$). For Ni$_{40}$Mo$_{40}$W$_{20}$, the machine learning model predicts $E\approx183$~GPa, where the refractory metals Mo and W are included [see Table~\ref{tab_3}]. The doping with refractory metals, nonmetals and lanthanides (e.g. B, Si, Gd, La) makes it possible to increase the strength of these alloys, which is actively used in modern metallurgy to produce heat-resistant alloys~\cite{Sanin_Kaplansky_2021,Wang_Shen_2016}. This simple quantitative analysis confirms that the properly selected composition and physical properties of the main chemical elements are most important in determining the alloys that best match the required mechanical properties.

\begin{table}[tbh]
\centering
\caption{Young's modulus $E$ predicted by machine learning model for binary and ternary amorphous metal alloys. Here, the average accuracy is $\sim87$~\%.}	
\begin{tabular}{lc|lc}
		\hline		\hline
	Alloy & $E$, GPa & Alloy & $E$, GPa \\
		\hline
	Cr$_{80}$B$_{20}$ &	$305$ & Cu$_{80}$Mg$_{20}$ & $60$ \\
	W$_{40}$Mo$_{40}$B$_{20}$ & $318$ & Cu$_{60}$Mo$_{40}$ & $154$ \\
	Ni$_{40}$Cr$_{40}$Co$_{20}$ & $58$ & W$_{40}$Ag$_{40}$B$_{20}$ & $234$ \\
	Ni$_{40}$Zr$_{40}$Si$_{20}$ & $108$	& Cr$_{40}$B$_{40}$Gd$_{20}$  & $217$ \\
	Ni$_{40}$Mo$_{40}$W$_{20}$ & $183$ & Cr$_{40}$Nb$_{40}$La$_{20}$ & $196$ \\
		\hline		\hline
\end{tabular}\label{tab_3}
\end{table}

\section{Conclusions}

In the present study, the machine learning model was applied to predict the Young's modulus $E$ and the yield strength $\sigma_{y}$ of amorphous metal alloys with different compositions. More than $50\,000$ different alloys were determined as well as $E$ and $\sigma_{y}$ were evaluated for each of them. It was found that the artificial neural network trained on the basis of information about the atomic number of a chemical element, its atomic mass, covalent radius, ionization energy, electronegativity, thermal conductivity, specific heat capacity, density, melting temperature and boiling temperature allows us to correctly determine of $E$ and $\sigma_{y}$ of amorphous metal alloys consisting $2$ to $7$ components and containing chemical elements with atomic numbers from $Z=3$ to $Z=79$. Here, the mean relative error is $\sim(12\pm1)$\% that is the good accuracy for the direct propagation multilayer neural network. The results of the statistical treatment made it possible to determine the chemical elements with the largest mass fraction, whose presence in the alloy leads to a significant increase in the strength of alloys. These chemical elements are B, Cr, Fe, Co, Ni, Nb, Mo, Pd and W. At the same time, the quantities $E$ and $\sigma_{y}$ show a weak dependence on the number of components in alloy. Thus, the most significant factors in the synthesis of alloys with the desired mechanical properties are the properly selected composition and the physical properties of the basic chemical element of alloy.

\section*{Acknowledgment}
This research was funded by the Russian Science Foundation (project no. 19-12-00022).

\end{document}